\documentclass[10pt,conference]{IEEEtran}
\usepackage{amsmath}
\usepackage{amssymb}
\usepackage{epic}

\newcommand{\PR}[1]{\mbox{{\bf P}$\left\{ #1 \right\}$}}
\newcommand{\EXP}[1]{\mbox{{\bf E}$\left\{ #1 \right\}$}}

\newtheorem{lemma}{Lemma}
\newtheorem{theorem}{Theorem}
\newtheorem{definition}{Definition}

\begin{document}

\author{\authorblockN{S\'andor Gy\H ori}\authorblockA{Department of Computer
Science and Information Theory\\ Budapest University of Technology
and Economics\\ H-1117, Magyar tud\'osok k\"or\'utja 2., Budapest,
Hungary\\ Email: gyori@szit.bme.hu}}
\title{Signature coding for OR channel\\with asynchronous access}
\maketitle

\def\thefootnote{}
\footnotetext{This work was sponsored by the Office of Naval
Research International Field Office and the Air Force Office of
Scientific Research, Air Force Material Command, USAF, under grant
number FA8655-05-1-3017. The U.S Government is authorized to
reproduce and distribute reprints for Governmental purpose
notwithstanding any copyright notation thereon.}

\begin{abstract}
Signature coding for multiple access OR channel is considered. We
prove that in block asynchronous case the upper bound on the
minimum code length asymptotically is the same as in the case of
synchronous access.
\end{abstract}

\section{Introduction}

Cohen, Heller and Viterbi \cite{CoHeVi71} introduced the model of
the noiseless OR channel for multiple access communication. If
there are $T$ users in the system and $x_i$ denotes the binary
message of the $i^{\text{th}}$ user, then the output of the
channel is defined by the Boolean sum of the message
\[
y=\bigvee\limits_{i=1}^T{x_i},
\]
i.e., the output is 0 iff all inputs are 0.

A possible example of communication scheme where this simple model
is suitable, is on/off keying (OOK) modulation. The bit 1
corresponds to a waveform and the bit 0 corresponds to the
waveform constant 0. The receiver consists of an envelope detector
followed by a threshold detector, so the demodulation is just a
decision whether all users sent the 0 waveform.

In multiple access information transmission three tasks should be
solved:
\begin{itemize}
\item identification of active users,
\item synchronization of their code words, and
\item decoding the messages.
\end{itemize}

In this article we are investigating the identification and
synchronization problem which is called \emph{signature coding} via a
multiple access OR channel. There are $T$ users in the
communication system, and each of them has only one own code word.
Becoming active a user sends his code word into the channel, and
otherwise does nothing, formally sends the all-zero code word into
the channel. The decoder, from the Boolean sum of the code words
of the active users, should reconstruct the set of active users.

For permanent activity of the users, the synchronous access is
trivial, with time sharing the maximum utilization 1 can be
achieved. For partial activity, however, the problem is hard and
is far from being solved. It is usually assumed that at most $M$
users communicate on the channel simultaneously, where $M$ is a
fixed number. The signature coding problem is to find a code of
minimum length $n_{\text{syn}}(T,M)$ such that if at most $M$
active out of $T$ total users send their code words, then from the
output vector of the OR channel the set of active users can be
identified.

For asynchronous access, the signature coding means the tasks of
identification and synchronization. In this paper we give upper
bound for $n_{\text{asyn}}(T,M)$, and show that the bounds for
synchronous and asynchronous access are asymptotically equal.

\section{Synchronous access}

There are a lot of study on the signature coding for multiple
access OR channel in the literature, but all of them assume
\emph{frame synchronous} access among users (all active users
begin transmitting their code words at the same time). Dyachkov
and Rykov \cite{DyRy83}, Erd\H os et al.\ \cite{ErFrFu85}, Hwang
and T.\ S\'os \cite{HwangTSos}, A and Zeisel \cite{AZeisel},
Ruszink\'o \cite{Ruszinko}, F\"uredi \cite{Furedi} gave lower and
upper bounds on the minimum code length $n$.

We say that a sequence $\mathbf{y}=(y_1,\dots,y_n)$ \emph{covers}
a sequence $\mathbf{z}=(z_1,\dots,z_n)$, i.e., $\mathbf{y}\ge\mathbf{z}$,
if $y_i\ge z_i,\; \forall i=1,2,\dots,n$.
Let
$\mathcal{C}=\{\mathbf{c}_1,\mathbf{c}_2,\dots,\mathbf{c}_T\}$ be
a code.
\begin{definition}
A code is Zero False Drop (ZFD) of order $M$, if every Boolean sum
of up to $M$ different code words covers no code word
other than those used to form the sum.
\end{definition}

This provides a really fast detection rule, i.e., if
$\mathbf{c}_i\le\mathbf{y}$, where $\mathbf{y}$ is the output
block of the channel, then the receiver declare user $i$ as
active. If the number of all potential users $T$ is known, and it
can be guaranteed that only a small fraction of them are active
simultaneously ($M\ll T$), then the ZFD code can be applied for
communicating via an OR channel with synchronous access.

In Theorem \ref{thm:DR} we repeat the result of Dyachkov and Rykov
\cite{DyRy83} on the upper bound on code length $n$ for the easy
comparability with the asynchronous access in Section
\ref{sec:async}. We denote by $\lesssim$ upper bound which hold
asymptotically in case of some given conditions.

\begin{theorem}[Dyachkov and Rykov \cite{DyRy83}]\label{thm:DR}
If $M$ is fixed, $T\to\infty$ and frame synchronous access is
used, then \vspace*{-1mm}
\[
n_{\text{syn}}(T,M)\lesssim\mathrm{e}\ln 2 \,(M+1)^2\log T.
\]
\end{theorem}

%\vspace*{2mm}
\begin{proof}
Consider a binary random code $\mathcal{C}$ of length $n$. In a
code word each bit is 1 with probability $p$ and 0 with
probability $1-p$, independently of each other, so the number of
1's in a code word is binomially distributed. $\mathcal{C}$ does
not meet the requirements if we can choose $M$ code words out of
the total number $T$ and another (tagged) code word such that for
each position where the tagged code word has 1's there is at least
one among the $M$ code words having also a bit 1. The probability
that in an arbitrary position the tagged user has a bit 1 while
all the active users have bit 0 is $p(1-p)^M$. Thus
\begin{equation}
\PR{\mbox{$\mathcal{C}$ is bad}}\le\binom{T}{M}(T-M)
\left(1-p(1-p)^M\right)^n.\label{eq:sync}
\end{equation}
This expression takes its minimum value at $p=\frac1{M+1}$, and
here
\begin{eqnarray}
\PR{\mbox{$\mathcal{C}$ is bad}}
&\le&\binom{T}{M}(T-M) \left(1-\tfrac1{M+1}\left(1-\tfrac1{M+1}\right)^M\right)^n\nonumber\\
&\le&T^{M+1} \left(1-\tfrac{\mathrm{e}^{-1}}{M+1}\right)^n\nonumber\\
&\le&T^{M+1} \mathrm{e}^{-\frac{n}{M+1}\mathrm{e}^{-1}}\nonumber\\
&=&\mathrm{e}^{(M+1)\ln 2\log
T-\frac{n}{M+1}\mathrm{e}^{-1}},\label{eq:syncexp}
\end{eqnarray}
where we applied that $\left(1-\frac1{M+1}\right)^M\ge
\mathrm{e}^{-1}$, and $1+x\le \mathrm{e}^x$ for all
$x\in\mathbb{R}$.

We need
\[
\PR{\mbox{$\mathcal{C}$ is bad}}<1,
\]
since then there is a good code. This gives an upper bound on the
minimum code length $n$. By taking the logarithm of
(\ref{eq:syncexp}) we get
\[
(M+1)\ln 2\log T-\frac{n}{M+1}\mathrm{e}^{-1}<0.
\]
The solution of this inequality is
\[
n>\mathrm{e}\ln 2 \,(M+1)^2\log T.
\]
If we choose the code length $n$ to
\[
n=(1+\delta)\,\mathrm{e}\ln 2 \,(M+1)^2\log T
\]
for an arbitrary constant $\delta>0$, the exponent in
(\ref{eq:syncexp}) becomes
\[
-\delta\ln2(M+1)\log T
\]
which tends to $-\infty$ when $T\to\infty$, that is why for such a
code length $n$
\[
\PR{\mbox{$\mathcal{C}$ is bad}}\to0.
\]
As the reasoning above is true for all arbitrarily small
$\delta>0$, the following asymptotic upper bound on the minimum
code length $n$ has been shown
\[
n_{\text{syn}}(T,M)\lesssim\mathrm{e}\ln 2 \,(M+1)^2\log T.
\]
\end{proof}

\section{Asynchronous access}\label{sec:async}

If frame asynchronous access is assumed, the coding method have to
ensure not just the identification but the synchronization, too.

\begin{theorem}\label{thm:asyn}
For frame asynchronous access, if $M$ is fixed and $T\to\infty$
\vspace*{-1.5mm}
\[
n_{\text{asyn}}(T,M)\lesssim\mathrm{e}\ln 2 \,(M+1)^2\log T.
\]
\end{theorem}

\vspace*{1.7mm} The detection is done by the following algorithm.
A sliding window is used which length equals to the code length
$n$. If, starting at a position, the binary vector of the channel
output covers the code word of a user, then it is declared as
active (\emph{identification}) beginning at this position
(\emph{synchronization}). Obviously, two different types of errors
can happen: false identification, and false synchronization.

During the design of the code it is supposed that the decoding
algorithm does \emph{not} have a memory (stateless). We have
synchronization error only when a code word is covered by the
\emph{beginning} of its shifted version and some other code words.
During the application of this code we use a decoding algorithm
with memory (stateful). If a user is declared as active beginning
at a given position, then he will be active in the next $n$ time
slots, so the algorithm need not to check its coverage in the next
$n$ time slots. Consequently, it does not cause synchronization
error if a code word is covered by the \emph{end} of its shifted
version and some other code words.

In the sequel it is supposed that \emph{exactly} $M$ users are
active simultaneously (in each time slot), which gives us an upper
bound on the covering probability compared to the original case,
when \emph{at most} $M$ users are active.

Identification error occurs if the Boolean sum of the active code
words covers the code word of another user.

\begin{lemma}
For frame asynchronous access, if $p=\frac1{M+1}$ \vspace*{-1mm}
\begin{equation}
\PR{\mbox{false identification}}\le \mathrm{e}^{(M+1)\ln T+M\ln
n-\frac{n}{M+1}\mathrm{e}^{-1}}.\label{eq:idexp}
\end{equation}
\end{lemma}

\vspace*{2mm}
\begin{proof}
As the bits of the code words of the users are chosen
independently of each other, the identification error probability
can be similarly calculated as in Theorem \ref{thm:DR}. Let us
select $M$ arbitrarily shifted code words, and another (tagged)
code word. The probability that in a given position the tagged
code word has an uncovered 1 is $p(1-p)^M$. That is why
\[
\PR{\mbox{false ident.}}\le\binom{T}{M}(T-M)n^M
\left(1-p(1-p)^M\right)^n,
\]
where the factor $n^M$ is needed because of the shift of the code
words. Let $p:=\frac1{M+1}$, then
\begin{eqnarray*}
\lefteqn{\PR{\mbox{false identification}}}\\
&&\le\binom{T}{M}(T-M)n^M\left(1-\tfrac1{M+1}\left(1-\tfrac1{M+1}\right)^M\right)^n\\
&&\le T^{M+1}n^M\left(1-\tfrac{\mathrm{e}^{-1}}{M+1}\right)^n\\
&&\le T^{M+1}n^M \mathrm{e}^{-\frac{n}{M+1}\mathrm{e}^{-1}}\\
&&=\mathrm{e}^{(M+1)\ln T+M\ln n-\frac{n}{M+1}\mathrm{e}^{-1}}.
\end{eqnarray*}
\end{proof}

\begin{figure*}[tb]
\centering \unitlength 0.55mm
\begin{picture}(270,40)(0,-5)
\put(0,0){\line(1,0){95}} \put(115,0){\line(1,0){80}}
\put(0,10){\line(1,0){95}} \put(115,10){\line(1,0){125}}
\put(0,20){\line(1,0){95}} \put(115,20){\line(1,0){125}}
\put(0,30){\line(1,0){95}} \put(115,30){\line(1,0){80}}
\multiput(0,0)(15,0){7}{\line(0,1){10}}
\multiput(120,0)(15,0){6}{\line(0,1){10}}
\multiput(45,10)(15,0){4}{\line(0,1){10}}
\multiput(120,10)(15,0){9}{\line(0,1){10}}
\multiput(0,20)(15,0){7}{\line(0,1){10}}
\multiput(120,20)(15,0){6}{\line(0,1){10}}
\put(173,4){\makebox(0,0){$c_k$}}
\put(173,14){\makebox(0,0){$c_{k-1}$}}
\put(128,4){\makebox(0,0){$c_{k-1}$}}
\put(128,14){\makebox(0,0){$c_{k-2}$}}
\put(68,4){\makebox(0,0){$c_2$}} \put(68,14){\makebox(0,0){$c_1$}}
\put(23,4){\makebox(0,0){$c_1$}}
\put(106,5){\makebox(0,0){$\cdots$}}
\put(106,15){\makebox(0,0){$\cdots$}}
\put(260,4){\makebox(0,0){tagged}}
\put(260,14){\makebox(0,0){shifted}}
\put(259,24){\makebox(0,0){others}} \put(2,13){\vector(1,0){43}}
\put(43,13){\vector(-1,0){43}} \put(22,17){\makebox(0,0){$d$}}
\put(23,24){\makebox(0,0){$U_1$}}
\put(68,24){\makebox(0,0){$U_2$}}
\put(128,24){\makebox(0,0){$U_{k-1}$}}
\put(173,24){\makebox(0,0){$U_k$}}
\put(23,-6){\makebox(0,0){$D_j$}}
\put(68,-6){\makebox(0,0){$D_j$}}
\put(128,-6){\makebox(0,0){$D_j$}}
\put(173,-6){\makebox(0,0){$D_j$}} \linethickness{0.5mm}
\put(0,0){\line(0,1){30}} \put(195,0){\line(0,1){30}}
\end{picture}
\caption{Bits of the tagged user and other active
users}\label{fig}
\end{figure*}

Synchronization error occurs if a code word is covered by the
shifted version of itself and some other active users. Depending
on the number of bits $d$ with which the code word of the tagged
user is shifted, disjoint classes of positions $D_1,\dots,D_d$ can
be distinguished, where
\[
D_j=\{j+\ell d:\; \ell=0,1,\dots,k-1 \; \mbox{ and }
k=\left\lfloor\tfrac{n-j}{d}\right\rfloor+1\}
\]
($j=1,\dots,d$). Each position belongs to exactly one class. All
classes have $k=\left\lfloor\frac nd\right\rfloor$ or
$\left\lceil\frac nd\right\rceil$ elements, and
$|D_1|+\cdots+|D_d|=n$.

The probability $f(D_j)$ that in an arbitrary class of positions
$D_j$ the tagged user has no uncovered 1's, can be derived in a
recursive way, starting at the last position. It is supposed that
$D_j$ contains $k=|D_j|$ positions. (We note that the probability
$f(D_j)$ depends only on the size of $D_j$ and not on the actual
elements of it.) For the sake of simplicity we use $1,2,\dots,k$
as position indices instead of $j,j+d,\dots,j+(k-1)d$. $c_\ell$
denotes the component of the code word of the tagged user at
position $\ell$ ($\ell=1,\dots,k$), and let $U_\ell$ be 0 if and
only if all the other users have 0 at this position (else it is
1). As the shifted code word of the tagged user is still not
active at the first position of the class $D_j$, there should be
considered $M$ users instead of $M-1$ in the calculation of $U_1$.
(Remember, that exactly $M$ active users were supposed in each
position.) That is why for all $\ell=1,\dots,k$
\[
\PR{c_\ell=0}=1-p,\qquad
\PR{c_\ell=1}=p,
\]
and
\begin{eqnarray*}
\PR{U_\ell=0}&=&\left\{\begin{array}{ll}(1-p)^{M-1}, & \mbox{if }
\ell=2,\dots,k,\\[1mm]
(1-p)^M, & \mbox{if } \ell=1 \end{array}\right.\\[1mm]
\PR{U_\ell=1}&=&1-\PR{U_\ell=0}
\end{eqnarray*}
(cf.\ Fig.\ \ref{fig}).

In the sequel we use the following
\begin{lemma}\label{lem:feltve}
If $V,W$ and $Z$ are independent random variables, and
$f(\cdot),g(\cdot)$ are arbitrary functions, then
\vspace*{-1mm}
\[
\EXP{f(V,W)g(V,Z)\mid V}=\EXP{f(V,W)\mid V}\EXP{g(V,Z)\mid V}.
\]
\end{lemma}

\vspace*{2mm}
\begin{proof}
\begin{eqnarray*}
\lefteqn{\EXP{f(V,W)g(V,Z)\mid V}}\\
&&=\EXP{\EXP{f(V,W)g(V,Z)\mid V,W}\mid V}\\
&&=\EXP{f(V,W)\EXP{g(V,Z)\mid V,W}\mid V}\\
&&=\EXP{f(V,W)\EXP{g(V,Z)\mid V}\mid V}\\
&&=\EXP{g(V,Z)\mid V}\EXP{f(V,W)\mid V}.
\end{eqnarray*}
\end{proof}

\begin{lemma}\label{lem:class}
For frame asynchronous access%, if $p=\frac1{M+1}$
\begin{eqnarray*}
\lefteqn{\PR{\mbox{\upshape code word of the tagged user is
covered}}}\\
&&\le\left(1-p(1-p)^M\right)^n.\hspace*{3cm}
\end{eqnarray*}
\end{lemma}

\vspace*{2mm}
\begin{proof}
Let us introduce the sequence of events
\begin{eqnarray*}
A_\ell&:=&\{\mbox{position $\ell$ is covered}\}\\
&=&\left\{\begin{array}{l}
\{c_{\ell-1}=1\}\cup\left\{\{c_{\ell-1}=0\}\cap\{c_\ell=1,U_\ell=0\}^c\right\},\\[1mm]
\quad\mbox{if } \ell=2,\dots,k,\\[2mm]
\{c_1=1,U_1=0\}^c,\;\mbox{ if } \ell=1,\end{array}\right.
\end{eqnarray*}
where $\{\;\}^c$ denotes the complement of an event. Thus
\[
f(D_j):=\PR{\mbox{all 1's in class $D_j$ are
covered}}=\PR{\bigcap\limits_{\ell=1}^k A_\ell}.
\]

We denote by $a_i^\phi\; (i=1,\dots,k,\;\phi=0,1)$ the conditional
probabilities that there is no uncovered 1 up to the
$i^{\text{th}}$ position given that the tagged user has a 0
($\phi=0$) or 1 ($\phi=1$) at the $i^{\text{th}}$ position
($c_i=0$ or 1), respectively.
\[
a_i^\phi:=\PR{\bigcap\limits_{\ell=1}^{i}A_\ell\mid\mbox{$c_i=\phi$}}.
\]
Hence
\begin{eqnarray*}
f(D_j)&=&\PR{\bigcap\limits_{\ell=1}^k A_\ell}\\
&=&\PR{\bigcap\limits_{\ell=1}^k A_\ell\mid c_k=1}\PR{c_k=1}\\
&&+\:\PR{\bigcap\limits_{\ell=1}^k A_\ell\mid c_k=0}\PR{c_k=0}\\
&=&pa_k^1+(1-p)a_k^0.
\end{eqnarray*}
Let us apply Lemma \ref{lem:feltve} with
$V=\{c_{i-1},c_i\},\;W=\{c_1,\dots,c_{i-2},U_1,\dots,U_{i-1}\},Z=\{U_i\}$,
and
$f(V,W)=I_{\left\{\bigcap\limits_{\ell=1}^{i-1}A_\ell\right\}},\;
g(V,Z)=I_{\{A_i\}}$. (Note, that $\PR{B}=\EXP{I_{\{B\}}}$ for an
arbitrary event $B$.)
\begin{eqnarray*}
\lefteqn{\PR{\bigcap\limits_{\ell=1}^{i-1}A_\ell\cap A_i\mid
c_i,c_{i-1}}}\\
&&=\EXP{I_{\left\{\bigcap\limits_{\ell=1}^{i-1}A_\ell\right\}}I_{\{A_i\}}\mid c_i,c_{i-1}}\\
&&=\EXP{f(V,W)g(V,Z)\mid V}\\
&&=\EXP{f(V,W)\mid V}\EXP{g(V,Z)\mid V}\\
&&=\EXP{I_{\left\{\bigcap\limits_{\ell=1}^{i-1}A_\ell\right\}}\mid c_i,c_{i-1}}\EXP{I_{\{A_i\}}\mid c_i,c_{i-1}}\\
&&=\PR{\bigcap\limits_{\ell=1}^{i-1}A_\ell\mid c_i,c_{i-1}}
\PR{A_i\mid c_i,c_{i-1}}
\end{eqnarray*}
By using this result we have for the conditional probabilities
\begin{eqnarray*}
a_i^\phi&=&\PR{\bigcap\limits_{\ell=1}^{i}A_\ell\mid c_i=\phi}\\
&=&\PR{\bigcap\limits_{\ell=1}^{i-1}A_\ell\cap
A_i\mid c_i=\phi}\\
&=&\sum\limits_{\psi=0}^1\PR{\bigcap\limits_{\ell=1}^{i-1}A_\ell\cap
A_i\mid c_i=\phi, c_{i-1}=\psi}\PR{c_{i-1}=\psi}\\
&=&\sum\limits_{\psi=0}^1\PR{\bigcap\limits_{\ell=1}^{i-1}A_\ell\mid
c_i=\phi, c_{i-1}=\psi}\\
&&\cdot\:\PR{A_i\mid c_i=\phi, c_{i-1}=\psi}\PR{c_{i-1}=\psi}\\
&=&\sum\limits_{\psi=0}^1a_{i-1}^\psi \PR{A_i\mid c_i=\phi,
c_{i-1}=\psi}\PR{c_{i-1}=\psi}
\end{eqnarray*}
($i\ge2$), thus
\begin{eqnarray*}
a_i^1&=&pa_{i-1}^1+(1-p)\left(1-(1-p)^{M-1}\right)a_{i-1}^0,\\
a_i^0&=&pa_{i-1}^1+(1-p)a_{i-1}^0.
\end{eqnarray*}

The first position of the class $D_j$ is different from the
others, because the shifted code word of the tagged user is not
active here.
\begin{eqnarray*}
a_1^\phi&:=&\PR{A_1\mid\mbox{$c_1=\phi$}}\\
&=&\PR{\{c_1=1,U_1=0\}^c\mid\mbox{$c_1=\phi$}}\\
&=&1-\PR{c_1=1,U_1=0\mid\mbox{$c_1=\phi$}}
\end{eqnarray*}
so
\begin{eqnarray*}
a_1^1&=&1-(1-p)^M,\\
a_1^0&=&1.
\end{eqnarray*}

Introduce the notation
\[
\mathbf{A}=\begin{bmatrix}
p, & (1-p)(1-(1-p)^{M-1})\\
p, & 1-p
\end{bmatrix},
\]
then
\begin{eqnarray*}
f(D_j)&=&\begin{bmatrix} p, & 1-p
\end{bmatrix}
\begin{bmatrix}
a_k^1\\a_k^0
\end{bmatrix}\\
&=&\begin{bmatrix} p, & 1-p
\end{bmatrix}
\mathbf{A}
\begin{bmatrix}
a_{k-1}^1\\a_{k-1}^0
\end{bmatrix}\\
&\vdots&\\
&=&\begin{bmatrix} p, & 1-p
\end{bmatrix}
\mathbf{A}^{k-1}
\begin{bmatrix}
a_1^1\\a_1^0
\end{bmatrix}\\
&=&\begin{bmatrix} p, & 1-p
\end{bmatrix}
\mathbf{A}^{k-1}
\begin{bmatrix}
1-(1-p)^M\\
1
\end{bmatrix}.
\end{eqnarray*}
For calculating the power of matrix $\mathbf{A}$ firstly its
diagonal form is needed. It has two eigenvalues
\begin{eqnarray*}
\lambda_1&=&\frac12+\frac12\sqrt{1-4p(1-p)^M},\\
\lambda_2&=&\frac12-\frac12\sqrt{1-4p(1-p)^M},
\end{eqnarray*}
and the corresponding eigenvectors are
\[
v_1=\begin{bmatrix}\frac{\lambda_1-1+p}{p}, &
1\end{bmatrix}^T,\qquad
v_2=\begin{bmatrix}\frac{\lambda_2-1+p}{p}, & 1\end{bmatrix}^T.
\]
Thus, the decomposition of matrix $\mathbf{A}$ is
\[
\mathbf{A}=\begin{bmatrix}\frac{\lambda_1-1+p}{p} &
\frac{\lambda_2-1+p}{p}\\[2mm] 1&1\end{bmatrix}
\begin{bmatrix}\lambda_1&0\\[2mm]0&\lambda_2\end{bmatrix}
\begin{bmatrix}\frac{p}{\lambda_1-\lambda_2}&
-\frac{\lambda_2-1+p}{\lambda_1-\lambda_2}\\[2mm]
-\frac{p}{\lambda_1-\lambda_2}&
\frac{\lambda_1-1+p}{\lambda_1-\lambda_2}
\end{bmatrix},
\]
the $(k-1)^{\text{th}}$ power of $\mathbf{A}$ is
\[
\mathbf{A}^{k-1}\!=\!\begin{bmatrix}\frac{\lambda_1-1+p}{p} &
\frac{\lambda_2-1+p}{p}\\[2mm] 1&1\end{bmatrix}
\!\!\begin{bmatrix}\lambda_1^{k-1}\!&0\\[2mm]0&\!\lambda_2^{k-1}\end{bmatrix}
\!\!\begin{bmatrix}\frac{p}{\lambda_1-\lambda_2}&
-\frac{\lambda_2-1+p}{\lambda_1-\lambda_2}\\[2mm]
-\frac{p}{\lambda_1-\lambda_2}&
\frac{\lambda_1-1+p}{\lambda_1-\lambda_2}
\end{bmatrix}\!,
\]
and the probability $f(D_j)$ is
\begin{eqnarray*}
\!\!\!\!&&f(D_j)=\left(1+\sqrt{1-4q}\right)^{\!k-2}\hspace*{-1pt}2^{-(k-2)}
\!\left(\hspace*{-1pt}\frac{\frac12-2q+q^2}{\sqrt{1-4q}}+\frac12-q\hspace*{-1pt}\right)\\
\!\!\!\!&&\;\;\;\;\;-\left(1-\sqrt{1-4q}\right)^{\!k-2}2^{-(k-2)}
\left(\frac{\frac12-2q+q^2}{\sqrt{1-4q}}-\frac12+q\right),
\end{eqnarray*}
where $q=p(1-p)^M$. Notice, that $0\le q\le\frac14$ for all
$M\ge1$ and $0\le p\le1$. By considering that
\[
\left(\frac{\frac12-2q+q^2}{\sqrt{1-4q}}-\frac12+q\right)\ge0,
\]
and
\begin{equation}
\left(\frac{\frac12-2q+q^2}{\sqrt{1-4q}}+\frac12-q\right)\le(1-q)^2\label{eq:0.228}
\end{equation}
(where (\ref{eq:0.228}) is true only for $q\le0.228$), $f(D_j)$
can be upper bounded
\begin{eqnarray*}
f(D_j)&\le&\left(1+\sqrt{1-4q}\right)^{k-2}2^{-(k-2)}(1-q)^2\\
&=&\left(\frac12+\sqrt{\frac14-q}\right)^{k-2}(1-q)^2\\
&\le&\left(1-q\right)^{k-2}(1-q)^2\\
&=&(1-q)^k\\
&=&\left(1-p(1-p)^M\right)^k.
\end{eqnarray*}

Remember, that we have applied (\ref{eq:0.228}) which is false for
$q>0.228$. This case can only happen if $M=1$, because if $M\ge2$,
then $q=p(1-p)^M\le\frac{4}{27}=0.148$. We need to consider the
case $M=1$. For $M=1$ the covering probability $f(D_j)$ can be
expressed in a rather simple form
\[
f(D_j)=\frac{(1-p)^{k+2}-p^{k+2}}{1-2p}.
\]
We show by induction that
\begin{equation}
\frac{(1-p)^{k+2}-p^{k+2}}{1-2p}\le\left(1-p(1-p)\right)^k.
\label{eq:TI0.228}
\end{equation}
For $k=1$ (\ref{eq:TI0.228}) is true. Let us suppose that
(\ref{eq:TI0.228}) is true for all $i\le k$, and it needs to be
true for $k+1$. If $p<\frac12$ we have to show it for $k+1$, iff
\begin{eqnarray*}
\frac{(1-p)^{k+3}-p^{k+3}}{(1-p)^{k+2}-p^{k+2}}&\stackrel{?}{\le}&
1-p(1-p)\\
(1-p)^{k+3}-p^{k+3}&\stackrel{?}{\le}&(1-p)^{k+2}-p^{k+2}-p(1-p)^{k+3}\\
&&+\:(1-p)p^{k+3}\\
0&\stackrel{?}{\le}&(1-p)^2p^2\left((1-p)^k-p^k\right)
\end{eqnarray*}
which is true for all $p<\frac12$. Inequality (\ref{eq:TI0.228})
can be similarly proven by induction for all $p>\frac12$.

\enlargethispage{-1mm} As the components of the code words are
chosen independently of each other, and classes $D_j$'s are
disjoint, we have
\begin{eqnarray*}
\lefteqn{\PR{\mbox{code word of the tagged user is covered}}}\\
&&=\PR{\bigcap\limits_{j=1}^d \{\mbox{all 1's in class $D_j$ are
covered}\}}\\
&&=\prod\limits_{j=1}^d\PR{\mbox{all 1's in class $D_j$ are
covered}}\\
&&=\prod\limits_{j=1}^d f(D_j)\\
&&\le\prod\limits_{j=1}^d \left(1-p(1-p)^M\right)^{|D_j|}\\[-1.5mm]
&&=\left(1-p(1-p)^M\right)^{\sum\limits_{j=1}^d|D_j|}\\
&&=\left(1-p(1-p)^M\right)^n,
\end{eqnarray*}
so the probability of false synchronization in the asynchronous
case is upper bounded by the probability of the false detection in
synchronous case.
\end{proof}

\begin{lemma}
For frame asynchronous access, if $p=\frac1{M+1}$ \vspace*{-1mm}
\begin{equation}
\PR{\mbox{false synchronization}}\le\mathrm{e}^{M\ln T+M\ln
n-\frac{n}{M+1}\mathrm{e}^{-1}}.\label{eq:asexp}
\end{equation}
\end{lemma}

\vspace*{2mm}
\begin{proof}
Let us select $M-1$ arbitrarily shifted code words, and another
(tagged) code word which is also active, but with some shift. By
Lemma \ref{lem:class} we have
\[
\PR{\mbox{false synch.}}\le\tbinom{T}{M-1}(T-M+1)n^M
\left(1-p(1-p)^M\right)^n\!\!,
\]
where the factor $n^M$ is needed because of the shift of the code
words. Let $p:=\frac1{M+1}$, then
\begin{eqnarray*}
\lefteqn{\PR{\mbox{false synchronization}}}\\
&&\le\tbinom{T}{M-1}(T-M+1)n^M\left(1-\tfrac1{M+1}\left(1-\tfrac1{M+1}\right)^M\right)^n\\%[-1.09mm]
&&\le T^M n^M\left(1-\tfrac{\mathrm{e}^{-1}}{M+1}\right)^n\\
&&\le T^M n^M\mathrm{e}^{-\frac{n}{M+1}\mathrm{e}^{-1}}\nonumber\\
&&= \mathrm{e}^{M\ln T+M\ln n-\frac{n}{M+1}\mathrm{e}^{-1}}.
\end{eqnarray*}
\end{proof}

\begin{proof}[Proof of Theorem \ref{thm:asyn}]
If a randomly chosen code $\mathcal{C}$ which has $T$ code words
of length $n$ satisfy the requirements of identification and
synchronization, then $\mathcal{C}$ can be applied for $T$ users
in communication via a multiple access OR channel. Obviously,
\[
\PR{\mbox{$\mathcal{C}$ is bad}}\le\PR{\mbox{false
ident.}}+\PR{\mbox{false synch.}}
\]
and we need
\[
\PR{\mbox{$\mathcal{C}$ is bad}}<1,
\]
since then there is a good code. This gives an upper bound on
minimum code length $n$. Thus, it is enough if the following
probabilities tend to 0
\begin{eqnarray}
\PR{\text{\upshape false identification}}&\to&0,\label{eq:id0}\\
\PR{\text{\upshape false synchronization}}&\to&0.\label{eq:syn0}
\end{eqnarray}
If we choose $p=\frac1{M+1}$, and the code length $n$ to
\[
n=(1+\delta)\,\mathrm{e}\ln 2 \,(M+1)^2\log T
\]
for an arbitrary constant $\delta>0$, the exponents in
(\ref{eq:idexp}) and (\ref{eq:asexp}) become
\begin{eqnarray*}
\lefteqn{-(M+1)\log T\Bigg(\delta\left(1-\tfrac\gamma{M+1}\right)\ln2}\\
&&\qquad\quad-\left(1-\tfrac1{M+1}\right)\frac{\ln\left((1+\delta)\,\mathrm{e}\ln2(M+1)^2\log
T\right)}{\log T}\Bigg),
\end{eqnarray*}
where constant $\gamma=0$ and 1, respectively. Both exponents tend
to $-\infty$ when $T\to\infty$, that is why we have (\ref{eq:id0})
and (\ref{eq:syn0}).

As the reasoning above is true for all arbitrarily small
$\delta>0$, the following asymptotic upper bound on the minimum
code length $n$ has been shown
\[
n_{\text{asyn}}(T,M)\lesssim\mathrm{e}\ln 2 \,(M+1)^2\log T.
\]

\end{proof}

\section{Conclusions}

We investigated the signature coding for multiple access OR
channel with asynchronous access. From Theorem \ref{thm:asyn} it
follows that the upper bound on the minimum code length via random
coding is the same as in the case of synchronous access.

\bibliographystyle{IEEEtran}
\bibliography{cdma,book3}

\end{document}